\documentclass{article}

\begin{document}

\def\aeta{a({\eta})}
\def\a2eta{a^2({\eta})}
\def\ateta{a^3({\eta})}
\def\aceta{a^4({\eta})}
\def\apeta{a^5({\eta})}
\def\aprimeeta{a'({\eta})}
\def\Bjke{B_{1(k)}({\eta})}
\def\Bdke{B_{2(k)}({\eta})}
\def\Bke{B_{(k)}({\eta})}
\def\Bprimeke{B'_{(k)}({\eta})}
\def\Bbiske{B''_{(k)}({\eta})}
\def\comc{\mbox{c.c.}}
\def\cae{c_1(\eta)}
\def\cbe{c_2(\eta)}
\def\deltae{\delta({\eta})}
\def\deltake{\delta_{(k)}({\eta})}
\def\detagmn{\delta g_{\mu\nu}}
\def\diag{\mathop{\rm diag}}
\def\dx#1{{\rm d}{#1}}
\def\deltaeke{\delta\epsilon_{(k)}({\eta})}
\def\deltake{\delta_{(k)}({\eta})}
\def\deltaekex{\delta_{(k)}({\eta},{\bf x})}
\def\deltaeex{\delta\epsilon({\eta},{\bf x})}	
\def\deltavtke{\delta\vartheta_{(k)}({\eta})}
\def\deltavtex{\delta\vartheta(\eta,{\bf x})}
\def\deltavte{\delta\vartheta({\eta})}
\def\ee{\epsilon({\eta})}
\def\eprimee{\epsilon'({\eta})}
\def\element33{~\frac{\sin^2(\sqrt{K}\chi)}{K}}
\def\fa{f_1}
\def\fb{f_2}
\def\ga{g_1}
\def\gb{g_2}
\def\Fae{F_1({\eta})}
\def\Fbe{F_2({\eta})}
\def\Fce{F_3({\eta})}
\def\factore{{\sf a}({\eta})}
\def\factorprimee{{\sf a}'({\eta})}
\def\factorke{{\sf a}^2({\eta})}
\def\factorxi{{\sf a}({\xi})}
\def\factor2xi{{\sf a}^2({\xi})}
\def\gmn{g_{mn}}
\def\Gjke{G_{1(k)}({\eta})}
\def\Gdke{G_{2(k)}({\eta})}
\def\lamjke{\lambda_{1(k)}({\eta})}
\def\lamdke{\lambda_{2(k)}({\eta})}
\def\lame{{\lambda(\eta)}}
\def\lamkbise{\lambda_{(k)}''({\eta})}
\def\lamkprimee{\lambda_{(k)}'({\eta})}
\def\lamk{{\lambda_{(k)}}}
\def\lamke{\lambda_{(k)}({\eta})}
\def\muke{\mu_{(k)}({\eta})}
\def\mukbise{\mu_{(k)}''({\eta})}
\def\mukprimee{\mu_{(k)}'({\eta})}
\def\pe{p({\eta})}
\def\pprimee{p'({\eta})}
\def\Qkscalar{Q_{(k)}({\bf x})}
\def\Qcmn{Q_{;m}^{\phantom{;m};n}}
\def\Qmn{Q_{m}^{\phantom{m}n}}
\def\Pmn{P_{m}^{\phantom{m}n}}
\def\p#1#2{\frac{\partial#1}{\partial#2}}
\def\dop#1#2{\frac{{\rm d\,}#1}{{\rm d\,}#2}}
\def\thetakaeta{\theta_{(k)}(\eta)}
\def\thetaka{\vartheta_{(k)}}
\def\He{H({\eta})}
\def\H2e{H^2({\eta})}
\def\Hprimee{H'({\eta})}
\def\Hbise{H''({\eta})}
\def\hdeltaex{\widehat{\delta}({\eta},{\bf x})}
\def\hdeltaxix{\widehat{\delta}({\xi},{\bf x})}
\def\hXex{\widehat{X}(\eta,{\bf x})}
\def\hXkex{\widehat{X}_{(k)}(\eta,{\bf x})}
\def\hXke{\widehat{X}_{(k)}({\eta})}
\def\Xex{X({\eta},{\bf x})}
\def\Xkex{X_{(k)}({\eta},{\bf x})}
\def\hdeltake{\widehat{\delta}_{(k)}(\eta)}
\def\hdeltakex{\widehat{\delta}_{(k)}(\eta,{\bf x})}
\def\hdeltakbise{\widehat{\delta}_{(k)}''(\eta)}
\def\hdeltakprimee{\widehat{\delta}_{(k)}'(\eta)}
\def\hdelta{\widehat{\delta}}
\def\hvtheta{\widehat{\vartheta}}
\def\hvthetak{\widehat{\vartheta}_{(k)}}
\def\hthetake{\widehat{\theta}_{(k)}{(\eta})}
\def\hdeltae{\widehat{\delta}({\eta})}
\def\hvthetae{\widehat{\vartheta}({\eta})}
\def\hdeltaxix{\widehat{\delta}({\xi},{\bf x})}
\def\hXe{\widehat{X}({\eta})}
\def\Xke{X_{(k)}(\eta)}
\def\sounde{{c_{\rm s}({\eta})}}
\def\soundke{{c_{\rm s}^2({\eta})}}
\def\sound2xi{{c_{\rm s}^2({\xi})}}
\def\soundprimee{{c_{\rm s}'({\eta})}}
\def\aaa{{\mathcal A}_{(k)}}
\def\hXk{\widehat{X}_{(k)}}
\def\hdeltavt{\widehat{\delta\vartheta}}
\def\hsfXex{\widehat{\sf X}(\eta,{\bf x})}
\def\hsfrex{\widehat{\sf r}(\eta,{\bf x})}
\def\hsfX{\widehat{\sf X}}
\def\hsfr{\widehat{\sf r}}
\def\hdeltavtke{\widehat{\delta\vartheta}_{(k)}({\eta})}
\def\cL{{\mathcal L}}
\def \Journal#1#2#3#4{{\em#1} {{\bf#2}{\rm,}} {#3} {(#4)}}
\def \AP{Adv.~Phys.}
\def \ApJ{Astrophys.~J.}
\def \JETP{Sov.~Phys.~JETP}
\def \MNRAS{Mon.~Not.~R.~Astron.~Soc.}
\def \ASS{Astroph.~Space Sci.}
\def \CQG{Class.~Quantum Grav.}
\def \JMP{J.~Math.~Phys.}
\def \PLA{{Phys.~Lett.}~A}
\def \PR{Phys.~Rep.}
\def \PRD{{Phys.~Rev.}~D}
\def \PRd{{Phys.~Rev.}}
\def \PRL{Phys.~Rev.~Lett.}

\title{Canonical gauge-invariant variables for scalar perturbations 
in synchronous coordinates}

\author{Zdzis{\l}aw A.~Golda, Andrzej Woszczyna and Karolina Zawada\\[10pt] 
 Astronomical Observatory, Jagellonian University\\  
ul. Orla 171, 30--244 Krak\'ow, Poland}

\date{} 
\maketitle
\begin{abstract}
Under an appropriate change of the perturbation variable 
Lifshitz-Khalat\-nikov propagation equations for the scalar perturbation reduce 
to d'Alembert equation. The change of variables is based on the Darboux 
transform.
\end{abstract} 


\section{Introduction\label{sec01}}

The gauge-invariant perturbation theories efficiently eliminate non-\break{}
physical perturbations, yet they provide the propagation equations in  
noncanonical form --- different for each theory. In this case, pure artefacts of
the choice of the reference system, gauge or the perturbation variable, are
likely to be confused with new dynamical phenomena yet unknown in the
laboratory-scale physics. This particularly concerns the large scale
limit, where some ``non-oscillatory  behaviour'' outside the particle horizon is
commonly expected.

In this paper we show that the definition of the gauge-invariant 
variables still possesses a freedom to chose the canonical variables, 
i.e.~variables which satisfy d'Alembert equation in its standard form. 
To show that we have used the classical Lifshitz-Khalatnikov formalism, 
which after being appropriately extended is equivalent to other ``manifestly'' 
gauge invariant descriptions. A~systematic construction of the 
gauge-invariant canonical variables is provided. Consequently, 
the scalar perturbation propagates like a massless scalar field 
in the Roberson-Walker space-time, and therefore there is a~close 
analogy between the perturbation theory (the acoustics of the 
expanding universes) and the  field theory in the curved space-time. 
Propagation of the sound waves in the early universe may be also 
considered as an example of the acoustic geometry in the sense of 
Unruh~\cite{Unruh1981&Unruh1995&Visser&Bilic&Golda&Woszczyna}.

The paper is organized as follows: In the second Section
we remind the method of elimination of spurious modes in the
Lifshitz-Khalatnikov theory~\cite{Lifshitz&Khalatnikov}. This Section is 
a~generalization of the procedure given in~\cite{Golda&WoszczynaJMP} to the case of
arbitrary space curvature and arbitrary density-dependent
pressure $p=p(\epsilon)$. In Section~\ref{sec03} we present
the general method of  construction of gauge-invariant
quantities in the synchronous system of reference.  We show that
the procedure is not unique, and therefore, there exist vast
classes of physically relevant gauge-invariant variables. 
The remaining freedom to choose between them may be used 
to better describe the perturbation dynamical properties, 
or to construct more adequate observables. A specific  
choice depends on the researcher motivation, 
and on the character of the problem to be solved. 
Finally in  Section~\ref{sec04} we adopt the previously 
defined methods to obtain a canonical form of the 
perturbation equations --- the d'Alembert equation.

\section{Synchronous system of reference --- techniques of
reduction of the gauge freedom}\label{sec02}

Consider the Robertson-Walker universe of arbitrary space
curvature ($K=-1,0,+1$) 
	\begin{equation}
g_{\mu\nu}=a^2(\eta)\,
	\diag
\left[-1,~1,~\element33,~\element33\sin^2\theta\right],
	\label{eq:geom_aku01}
	\end{equation}		
with the hydrodynamic energy-momentum tensor 
	\begin{equation}
T^{\mu\nu}=\left(\epsilon +p\right)u^\mu u^\nu+p\,g^{\mu\nu}
	\label{eq:tensor_cieczy}
	\end{equation}	
and the barotropic equation of state $p=p(\epsilon)$. We
investigate small perturbations $\delta g_{\mu\nu}$ of the
metric~(\ref{eq:geom_aku01}). We limit ourselves to scalar (density)
perturbations, therefore, $\delta g_{\mu\nu}$  is defined by two
scalar functions~$\lamke$ and~$\muke$
	\begin{eqnarray}
\delta g_{\mu0}&=&0,\\
\delta g_{m}^{\phantom{m}n}&=&\sum\limits_k\left(\lamke\Pmn
+\muke\Qmn\right)+{\comc},
	\label{eq:zaburzenie_metryki}
	\end{eqnarray}	
where $\Pmn$ and $\Qmn$ stand for scalar
harmonics~\cite{Lifshitz&Khalatnikov},   
	\begin{eqnarray}
\Qmn&=&\frac{1}{3}\Qkscalar\delta_m^{\phantom{m}n},\\
\Pmn&=&\frac{1}{k^2-K}\Qcmn+\Qmn,
	\label{eq:harmoniki}
	\end{eqnarray}	
$\Qkscalar$ is the complex solution of the Helmholtz equation
$\Qkscalar_{;m}^{\phantom{;m};m}=-(k^2-K)\Qkscalar$, ${\bf
x}=\{\chi,\theta,\varphi\}$, and the amplitudes
$\lamke$ and $\muke$ satisfy the system of two
second order equations convoluted to each other 
	\begin{eqnarray}
	\label{eq:lambda}
\lamkbise+2\frac{\aprimeeta}{\aeta}\lamkprimee-\frac{k^2-K}{3}\left[\lamke+\muke\right]&=&0,\\
\mukbise+\left[2+3\soundke\right]\frac{\aprimeeta}{\aeta}\mukprimee\hspace*{30mm}&&\nonumber\\
+\frac{k^2-4K}{3}\left[1+3\soundke\right]\left[\lamke+\muke\right]&=&0.
	\label{eq:mu}
	\end{eqnarray}	
$\sounde$ denotes the sound velocity: 
$\sounde=\sqrt{\pprimee\slash\eprimee}\neq0$. Then the density contrast\footnote{In 
more rigorous notation one should write $\delta=\delta\epsilon\slash\epsilon_0$, 
where $\epsilon_0$ denotes background, unperturbed energy density. For the sake 
of simplicity we skip the subscript ``0'' in formulas.}
$\delta=\delta\epsilon\slash\epsilon$ 
 is a~linear combination~\cite{Lifshitz&Khalatnikov} 
	\begin{equation}
\delta(\eta, {\bf x})=\sum\limits_k\deltake\,\Qkscalar+\comc, 
   	\label{eq:delta}
	\end{equation}
and
	\begin{equation}
\deltake=\frac{\aaa}{3\ee\a2eta}\left[(k^2-4K)(\lamke
+\muke)+3\frac{\aprimeeta}{\aeta}\mukprimee\right], 
   	\label{eq:deltak}
	\end{equation}
where $\aaa$ are arbitrary complex numbers. The system (\ref{eq:lambda}--\ref{eq:mu}) 
defines the four
dimensional phase-space, which means that the system possesses two
physical and two gauge degrees of freedom. Pure gauge solutions
are known, and for~$\lamke$ they respectively read~\cite{Lifshitz&Khalatnikov}
	\begin{eqnarray}
	\label{eq:fic01}
G_1(\eta) &=& 1,\\
G_2(\eta) &=& -(k^2-K)\int\frac{1}{\aeta}\dx{\eta}.
    \label{eq:fic02}
	\end{eqnarray}
Difficulties with fixing the gauge freedom have stimulated
elaboration of the gauge-invariant 
theories~\cite{Olson,Woszczyna&Kulak,Bruni&Dunsby&Ellis,Lyth&Stewart,Brandenberger&Kahn&Press,Mukhanov&Feldman&Brandenberger}. 
After the famous 
Bardeen's paper~\cite{Bardeen} these theories were intensively developed
for almost two decades. Yet, the original Lifshitz-Khalatnikov
formalism, when appropriately extended, provides an equally good
description of inhomogeneities. The procedure is as follows: we
reduce (\ref{eq:lambda}--\ref{eq:mu}) system to the single
fourth-order equation for $\lamke$.
	\begin{eqnarray}
&&\hspace*{-4mm}\lamk^{(4)}(\eta)+\aeta\He\left[4+3\soundke\right]\lamk^{(3)}(\eta)\nonumber\\
&&{}+\left\{-5K+\a2eta\left[2\left(\frac{\ee}{3}-\pe\right)+\left(9\H2e-\ee\right)\soundke\right]\right.\nonumber\\
&&{}+\left.\left(k^2-K\right)\soundke\phantom{\frac{|}{|}}\hspace*{-.4em}\right\}\lamkbise
+\aeta\He\left[-\a2eta\left[\ee+3\pe\right ]\right.\nonumber\\
&&{}+\left.2\a2eta\H2e\left[1+3\soundke\right]\right.+\left.\left(k^2-K\right)\soundke\right]\lamkprimee=0,
   	\label{eq:l4}
	\end{eqnarray}
where $\He$ stands for the Hubble parameter
$\He=\aprimeeta\slash \a2eta$. The knowledge of the gauge
solutions (\ref{eq:fic01}--\ref{eq:fic02}) enables one to
extract the gauge space from the space of all solutions. 
First, we write the solutions in
the form
	\begin{equation}\label{eq:lfull}
\lamke = 
	\fa\lamjke + \fb\lamdke + 
	\ga\Gjke + \gb\Gdke
   	\end{equation}	
with explicitly separated linear subspace $\ga\Gjke + \gb
\Gdke$ carrying all the gauge freedom ($\ga,\gb$ two
arbitrary coefficients). Subsequently, we adopt the Darboux
transform\footnote{Darboux transform of a~function $f(x)$ is defined  
as $\widehat{f}(x)=A(x)f(x)+B(x)f'(x)$ (compare also formula 
(\ref{eq:Darboux_tran}) in section~3) with arbitrary 
$x$-dependent coefficients~$A$ and~$B$. The same transform  
is extensively exploited in the soliton theory~\cite{Lamb}.}~\cite{Darboux} 
of $\lamke$ to express the two remaining (physical)
degrees of freedom. A new perturbation variable $\Bke$ appears
	\begin{equation}
\Bke=\aeta\dop{}{\eta}\!\!\left[\!\left({\dop{}{\eta}
\frac{\lamke}{\Gjke}}\right)\!\!\left({\dop{}{\eta}\frac{\Gdke}{\Gjke}}\right)^{{\!\!}-1}\right]=
\aeta\dop{}{\eta}\frac{\dop{}{\eta}
\lamke}{\dop{}{\eta}\Gdke}.
	\label{eq:Darboux}
	\end{equation}
On strength of (\ref{eq:l4}) the variable $\Bke$ satisfies the
second order equation
	\begin{eqnarray}
\Bbiske-\left[1-3\soundke\right]\aeta\He\,\Bprimeke\hspace*{30mm}\nonumber\\
-\left[\left(\frac13+\soundke\right)\a2eta \ee-\left(k^2-K\right)\soundke
\right]\Bke=0.
	\label{eq:giB}
	\end{eqnarray}
Equation (\ref{eq:giB}) does not contain gauge the coefficients, therefore,
the variable $\Bke$ is gauge-invariant. Now, the function
$\lamke$ can be found by the inverse Darboux transform of
the two linearly independent solutions to 
equation~(\ref{eq:giB})\footnote{The integration constants 
in the inverse Darboux transform can be set arbitrary, because
the gauge freedom is already guaranteed by free choice of
the coefficients~$\ga$ and~$\gb$.}
	\begin{eqnarray}
\lamke=f_1(K-k^2)\int\frac{1}{\aeta}\left[\int\frac{\Bjke}{\aeta}\dx{\eta}\right]\dx{\eta}\hspace{32mm}\nonumber\\
+f_2(K-k^2)\int\frac{1}{\aeta}\left[\int\frac{\Bdke}{\aeta}\dx{\eta}\right]\dx{\eta}
+\ga \Gjke + \gb \Gdke.
	\label{eq:lk}
	\end{eqnarray}
On can insert $\lamke$ to 
(\ref{eq:lambda}) and solve it algebraically to
obtain the second unknown function $\muke$. As 
a~result, by fixing four numbers $\{\!\fa,\fb,\ga,\gb\!\}$ one
uniquely determines the metric correction $\detagmn$. The metric
correction is not gauge-invariant and cannot be such in any
perturbation formalism. Quantities containing time integrals 
of~$\Bke$ would exhibit similar properties, while those
containing $\Bke$ and derivatives are obviously
gauge-invariant.

\section{Families of gauge-invariant variables in the
synchronous reference system\label{sec03}} 

The metric perturbation can be written in the form~(\ref{eq:lk})
if the solutions of the equation~(\ref{eq:giB}) are explicitly known.
However, even in the case when these solutions are not known one
may employ the equation~(\ref{eq:giB}) to ``convert''  some
non-gauge-invariant perturbation variables into gauge-invariant
ones. This particularly refers to perturbation measures based
on hydrodynamic quantities, since they involve the metric 
derivatives and are free of the metric elements themselves.
The procedure is following:\\
1) For the non-gauge-invariant perturbation variable $X$ we take the
linear combination of~$X$ and its time derivative $X'$ (the Darboux
transform)
	\begin{equation}
\hXex=\cae\left[\p{}{\eta}\Xex+\cbe\Xex\right], 
	\label{eq:Darboux_tran}
	\end{equation}
which means that 
	\begin{equation}
\hXex=\sum\limits_k\hXke\,\Qkscalar+\comc, 
	\label{eq:19}
	\end{equation}
and 
	\begin{equation}
\hXke=\cae\left[\dop{}{\eta}\Xke+\cbe\Xke\right]. 
	\end{equation}
2) We apply the equation (\ref{eq:giB}) to formally express
 the coefficients $\hXke$ in the form
	\begin{equation}
\hXke=\Fae\dop{}{\eta}\Bke+\Fbe\Bke+\Fce\int\frac{\Bke}{\aeta}\,\dx{\eta}.
	\label{eq:Darboux_tran_Lif}
	\end{equation}
3) We look for $c_1(\eta)$ and $c_2(\eta)$ such that 
$\hXke$ contains solely function $\Bke$ and its derivatives but
not integrals. (Integral over~$\eta$ in (\ref{eq:Darboux_tran_Lif}) 
introduces arbitrary constant~$c_{(k)}$, which depends on the wave 
vector~$\mbox{\bf k}$. The variable~$\hXex$ (formula~(\ref{eq:19}))
would contain then terms being arbitrary functions of the space coordinate \mbox{\bf x}.
In this way integration constant $c_{(k)}$ restores one degree of the gauge freedom.) 
The relevant condition $\Fce=0$ leads to
a~linear equation for $\cbe$, and leaves $\cae$ free.
Therefore, the new gauge-invariant variable $\hXex$ is
actually a~$\cae$-dependent family of variables, 
where $\cae$ is an arbitrary function of time. 
Below we show some examples.

\subsection{The density contrast}

On strength of (\ref{eq:delta}), (\ref{eq:lambda}--\ref{eq:mu}) and (\ref{eq:lk}) the density
contrast $\deltake$ can be expressed by use of the gauge
invariant function $\Bke$.
	\begin{eqnarray}
\deltake&=&{}-3\frac{\He}{\ateta\ee}\Bprimeke
-\frac{1}{\a2eta}\left[\frac{k^2-K}{\a2eta\ee}-1\right]\Bke\nonumber\\
&&{}+\frac{3}{2}\left[\pe+\ee\right]\frac{\He}{\ee}\int\frac{\Bke}{\aeta}\dx{\eta}.
	\label{eq:deltaB}
	\end{eqnarray}
The contrast $\deltake$ is not gauge-invariant, as it is well 
known~\cite{Bardeen}. The gauge freedom is caused
by the last term in (\ref{eq:deltaB}) containing the integral of $\Bke$.
Let us introduce a~new variable --- the ``modified density
contrast''
	\begin{equation}
	\label{eq:Darboux_tran_delta}
\hdeltake=\cae\left[\dop{}{\eta}\deltake+\cbe\deltake\right].
	\end{equation}
With the aid of (\ref{eq:giB}) one obtains 
	\begin{equation}
	\label{eq:Darboux_tran_delta_f3}
\hdeltake = \Fae\dop{}{\eta}\Bke + \Fbe\Bke + \Fce\int\frac{\Bke}{\aeta}\dx{\eta},
	\end{equation}
where the function  $\Fce$ reads\footnote{$\Fae$ and $\Fbe$
are irrelevant to the problem discussed here, therefore we skip
writing them explicitly.}
	\begin{equation}
\Fce=\frac{3}{2}c_1(\eta)\He\left[1+\frac{\pe}{\ee}\right] 
\left[\dop{}{\eta}\ln\left(\frac{1}{\aeta}\frac{\eprimee}{\ee}\right)+c_2(\eta)\right].
	\label{eq:f3}
	\end{equation}
By setting
	\begin{equation}
\cbe=-\dop{}{\eta}\ln\left(\frac{1}{\aeta}\frac{\eprimee}{\ee}\right),
	\label{eq:c2}
	\end{equation}
one eliminates the integral of $\Bke$ ($\Fce=0$), hence $\hdeltake$ 
becomes a~gauge independent quantity
	\begin{equation}
\hdeltake=c_1(\eta)\frac{\eprimee}{\aeta\ee}\dop{}{\eta}\left[\frac{\aeta\ee}{\eprimee}\deltake\right].
	\label{eq:hdeltake}
	\end{equation}
This variable is still defined up to the factor $c_1(\eta)$
being an arbitrary function of time.

\subsection{The expansion rate contrast}

Identically we proceed with inhomogeneities in the expansion
rate\footnote{$\vartheta\equiv u^\mu_{\phantom{\mu};\mu}$ is the expansion~\cite{Hawking&Ellis}.} 
$\delta\vartheta(\eta,{\bf x})$.
	\begin{equation}
\delta\vartheta(\eta,{\bf x})=\sum\limits_k\deltavtke\,\Qkscalar+\comc,
	\label{eq:theta}
	\end{equation}
where
	\begin{eqnarray}
\deltavtke&=&\frac{1}{\ateta}\left[\frac{k^2-K}{\a2eta(\pe+\ee)}-\frac32\right]\Bprimeke\nonumber\\
&&{}-\frac{\He}{\a2eta}\left[\frac{k^2-K}{\a2eta(\pe+\ee )}-\frac32\right] \Bke \nonumber\\
&&{}+\frac{3}{2}\left[\frac12 \left(\pe+\ee\right)-\frac{K}{\a2eta}\right]\int\frac{\Bke}{\aeta}\dx{\eta}\,.
	\label{eq:thetake}
	\end{eqnarray}
We introduce $\hdeltavtke$ defined as a linear combination 
	\begin{equation}
\hdeltavtke=\cae\left[\dop{}{\eta}\deltavtke+\cbe\deltavtke\right],
	\label{eq:thetakaeta}
	\end{equation}
and eliminate the integral of $\Bke$. We finally obtain
	\begin{equation}
\hdeltavtke=\cae\left[\dop{}{\eta}\deltavtke-\dop{}{\eta}\ln\left[\frac{\Hprimee}{\aeta}\right]\deltavtke\right].
	\label{eq:thetakaeta01}
	\end{equation}

\subsection{Other variables}

By employing the same procedure one may define gauge-invariant
variables which combine inhomogeneities in Hubble flow with
inhomogeneities of the energy density
	\begin{equation}
\hsfrex=\cae\left[\deltaeex-\frac{1}{3}\frac{\eprimee}{\Hprimee}\deltavtex\right].
	\label{eq:hXke}
	\end{equation}
It is clear that functions $F(\hdelta,\hdeltavt,\hsfr,\ldots,\epsilon,H,\ldots)$  of the
gauge-invariant variables $\{\hdelta,\hdeltavt,\hsfr,\ldots\}$ and the parameters of the
background universe  $\{\epsilon,H,\ldots\}$ are also gauge-invariant. None 
of these variables is preferred by the perturbation theory
itself. The choice of the variable depends on its physical
meaning and usefulness, its relation to other physical concepts and
theories, and eventually on the ability to construct relevant
observables.

\section{Perturbation equations in the canonical form\label{sec04}}

In all the examples above we constructed the gauge-invariant
variables which refer to the synchronous system of
reference. The function $\cbe$ was uniquely determined for
each $\hdeltae$, $\hvthetae$ and $\hXe$ by the demand of their gauge-invariance. The
freedom to make an arbitrary choice of the time-dependent factor $\cae$ still
remains, and therefore, we are now able to construct 
gauge-invariant variables with particular dynamical properties.
Consider the variable $\hdeltae$ with $\cae$ chosen as
	\begin{equation}
\cae=\a2eta\H2e\frac{\ee}{\eprimee}.
	\label{eq:c1eta}
	\end{equation}
We obtain the gauge-invariant variable\footnote{Please note the misprint 
in paper~\cite{Golda&WoszczynaPLA} formula~(7), already corrected in gr-qc\slash0302091.}
	\begin{equation}
\hdeltake=\aeta\H2e\dop{}{\eta}\left[\aeta\frac{\ee}{\eprimee}\deltake\right],
	\end{equation}
which satisfies the propagation equation of the form
	\begin{equation}
\hdeltakbise+\left[2\frac{\factorprimee}{\factore}-\frac{\soundprimee}{\sounde}\right]\hdeltakprimee+\soundke(k^2-K)\hdeltake=0,
	\label{eq:jawna}
	\end{equation}
with the function $\factore$ defined as
	\begin{equation}
\factore=\aeta\sqrt{\frac{1}{\sounde}\frac{\ee+\pe}{3\H2e}}\,.
	\end{equation}
Now we introduce the new time parameter $\dx{\xi}=\sounde\dx{\eta}$ (acoustic conformal 
time~\cite{Golda&WoszczynaPLA}). After this repar\-ametrization the propagation 
equation reduces to d'Alembert equation
	\begin{equation}
{\nabla}^\mu{\nabla}\!_\mu\,\hdeltaxix=0
	\label{eq:falowe}
	\end{equation}
in the Robertson-Walker space-time with coordinates
$\{\xi,\chi,\theta,\varphi\}$ and the metric form
	\begin{equation}
{\sf g}_{\mu\nu}=\factor2xi\,
	\diag
\left[-1,~1,~\element33,~\element33\sin^2\theta\right],
	\label{eq:geom_aku03}
	\end{equation}	
Therefore, the modified density contrast $\hdeltaxix$ propagates in the
Robertson-Walker space-time with the scale factor $\factorxi$ in
the same manner as minimally coupled scalar field in the
Robertson-Walker space-time with the scale factor $\aeta$~\cite{Birrell&Davies}. 
All the classical results obtained in the field
theory on the curved space-times apply to the density
perturbations in the expanding universe.
One can introduce the Lagrangian 
	\begin{equation}
\cL=\frac12{\sf g}^{\mu\nu}\widehat{\delta}_{,\mu}\widehat{\delta}_{,\nu}
	\end{equation}	
for field $\hdeltaxix$ and derive the equation of motion~(\ref{eq:falowe}) 
by use of the Euler-Lagrange equations 
	\begin{equation}
\nabla\!_{\mu}\p{}{\widehat{\delta}_{,\mu}}\cL-\p{}{\widehat{\delta}}\cL=0.
	\end{equation}	

An appropriate change of the perturbation variables~\cite{Golda&WoszczynaPLA} reduces 
the propagation equation of any gauge-invariant theory to the form 
of the equation~(\ref{eq:giB}). By use of the procedure discussed above 
one can construct canonical variables, and as a  consequence, 
each of these formalism can be expressed in the Lagrange-Hamilton 
language.
\medskip\smallskip

The authors thank L.M. Soko{\l}owski for reading the manuscript 
and for constructive critics.  

\bibliographystyle{plain}

\end{document}